\documentclass[intlimits,twoside,a4paper]{article}

\usepackage[cp1251]{inputenc}
\usepackage{multirow}

\usepackage[eqsecnum]{cmpj3}

\newcommand{\be}{\begin{equation}}
\newcommand{\ee}{\end{equation}}
\newcommand{\ba}{\begin{eqnarray}}
\newcommand{\ea}{\end{eqnarray}}

\issue{2017}{20}{4}{43703}
\doinumber{10.5488/CMP.20.43703}
\title[Estimation of electron temperature in heated metallic nanoparticle]%
{Estimation of electron temperature in heated metallic nanoparticle%
}
\author[N.I. Grigorchuk]{N.I. Grigorchuk}
\address{
Bogolyubov Institute for
Theoretical Physics of the National Academy of Sciences of Ukraine, \\
 14-b Metrologichna St., 03680 Kyiv, Ukraine 
}

\date{Received May 12, 2017, in final form June 19, 2017}
\begin{document}

\maketitle

\begin{abstract}
A method is proposed for determining the temperature of hot electrons in a
metallic nanoparticle embedded in a dielectric matrix under ultrashort laser pulses
irradiation. The amplitude and power of the longitudinal spherical acoustic
oscillations as functions of density and elastic properties of the medium,
the laser pulse duration, the electron temperature, radii of particles, and the electron-phonon
coupling constant are obtained. The efficiency of the electron energy transfer from
heated noble nanoparticles to a surrounding environment is estimated for different electron
temperatures.
\pacs 78.67.Bf, 68.49.Jk, 73.23.-b, 52.25.Os, 36.40.Vz
\end{abstract}

\section{Introduction}
Metallic nanoparticles (MNs) are mainly studied due to their unique optical
properties \cite{BH} and extensive practical applications \cite{CMF,KMH,TKC,DVC}.
Over the last decade, the acoustic oscillations of MNs excited by ultrashort
laser pulses have been under intense study \cite{BH,CMF,KMH,TKC,DVC,KNG,GT,G,JT}.
This is due to the availability of  important data relating to the elastic properties
of these particles and their mechanical coupling to the surrounding medium, which forms
the foundation for the design of elasticity sensors in the nanometer range \cite{AW}.
The generation of sound waves by MNs was originally observed experimentally for noble
nanoparticles~\cite{CMF,KMH,TKC}.

When MNs are excited by ultrashort pulses, the energy pulse is initially transferred
to the gas of free electrons which collide with one another as well as with the lattice
vibrations, and redistribute this energy (being thermalized) over a short time (of the
order of tens of picoseconds in a 100~{\AA} particle \cite{KMH}). The electron gas in
the particle is immediately heated. Due to a low heat capacity of the electron
gas (compared to the lattice), there originates a short (but strong) electron pressure
burst \cite{PGM}. It is equivalent to a short mechanical impact upon the particle
surface over an infinitely small time interval and can generate spherical sound
waves in the medium surrounding the particle \cite{VCL,PKL}.

In this article, we proposed a method for determining the temperature of hot electron
gas in a metallic nanoparticle embedded in a dielectric matrix under ultrashort laser pulses
irradiation. To accomplish this, the amplitude and power of the longitudinal sound wave
generated by the excess pressure of an electron gas inside the MN driven
by ultrashort laser pulses are calculated. So far, this problem has been little studied
theoretically.

\section{Initial principles and model}
Let an ultrashort laser pulse of duration $\tau_0$ be incident on a spherical MN
embedded in an infinite, elastically isotropic, dielectric medium. Energy transfer
from the laser to the lattice of a nanoparticle depends much on the
relationship between the particle sizes, the electron mean free path in the
particle, and the Debye length $l_{\text D} = \piup\upsilon_{\text F}/\omega_{\text D}$, which plays an
important role in the energy exchange between the hot electrons and the lattice \cite{Har}.
Its values for several metals are listed below in  table~\ref{tab2}.
We shall consider here the particles with radii $R < l_{\text D}/2$.

The additional pressure that develops during heating of the electron gas by the laser
pulse can produce oscillations of the nanoparticle surface, which in turn, generate
 longitudinal acoustic waves in the elastically isotropic space around the particle.

The equation for the propagation of longitudinal
acoustic oscillations in this space is given by \cite{LL}
\be
 \label{eq 1}
  \nabla^2 {\bf u}_{\text L}(t)-\frac{1}{s^2_{\text L}}\frac{\partial^2}{\partial t^2}{\bf u}_{\text L}(t)=0,
   \ee
where $s_{\text L}$ is the longitudinal sound speed in the medium. Elastic displacement waves
${\bf u}_{\text L}$ are accompanied by bulk compressions and expansions of the surrounding medium.

As boundary conditions for equation~(\ref{eq 1}), we take a condition that all forces applied to the
particle surface ($r = R$) are equal to zero. This equality can be presented as \cite{LL}
\be
 \label{eq 2}
  \left.\left[\sigma_{rr}-\frac{2{\cal{E}}_{\text s}}{R^2}u_r(t)\right]\right|_{r=R}=-\delta p(t).
   \ee
Here, $u_r = |{\bf u}_r|$ is the radial component of the displacement, $R$ is the particle
radius, ${\cal{E}}_{\text s}$ is the surface energy density, $\delta p(t)$ is the time dependent
additional pressure of the hot electrons, and $\sigma_{rr}$ is the stress tensor,
whose components are expressed through the moduli $K$, $\mu$ of the uniform
compression and rigidity, which one can find elsewhere (e.g., in  \cite{LL}).
For spherical acoustic oscillations, one can assume that $u_{\text L}(t) = u_r(t)$.

The pressure of the electron gas on the surface of
the nanoparticle can be written as the sum
\be
 \label{eq 3}
  p(t) = p_0+\delta p(t).
   \ee
Here, the first term represents the pressure of a degenerate Fermi gas at $T_{\text e} = 0$~K,
$p_0 = \frac{2}{5}n\mu_0$, where $n = N/V$ is the electron density, $V$ is the particle volume,
and $\mu_0$ is the limiting value of the chemical potential at $T_{\text e} = 0$~K ($\mu_0 = \varepsilon_{\text F}$,
where $\varepsilon_{\text F}$ is the Fermi energy). The second term, $\delta p(t)$, represents
the additional time dependent gas pressure owing to the electron mobility
at temperatures $T_{\text e} > 0$. It can be written as \cite{RR}
\be
 \label{eq 4}
  \delta p(t) = \alpha T^2_{\text e}(t), \qquad  \alpha = \frac{\piup^2}{6}n\frac{k^2_{\text B}}{\mu_0}\,.
    \ee

\section{Amplitude of the sound wave}
To solve equation~(\ref{eq 1}) with the boundary condition (\ref{eq 2}), we use the
potential method \cite{LL} ${\bf u}({\bf r},t)\equiv\pmb\nabla\varphi({\bf r},t)$.
Then, representing of function $\varphi({\bf r},t)$ in the form of spherical waves
expanding from particle center $\varphi({\bf r},t) = \phi\big(t-(r-R)/s_{\text L}\big)/r$,
where $\phi$ is an arbitrary, twice differentiable function and using the boundary
condition (\ref{eq 2}) at the nanoparticle surface, we found for $\phi$
an inhomogeneous differential equation
\be
 \label{eq 5}
  \frac{\partial^2}{\partial t^2}\phi(t)+2\gamma\frac{\partial}{\partial t}\phi(t)+
   \omega^2_0 \, \phi(t) = -\delta p(t)\frac{R}{\rho}\,,
    \ee
in which $\rho$ is the mass density of the medium surrounding the nanoparticle, and
\be
 \label{eq 6}
  \gamma = \frac{1}{s_{\text L} R}\left(2s^2_{\text T}+\frac{{\cal{E}}_{\text s}}{\rho R}\right), \qquad
   \omega^2_0 = \frac{2}{R^2}\left(2s^2_{\text T}+\frac{{\cal{E}}_{\text s}}{\rho R}\right).
    \ee
Here, $s_{\text T}$ is a transverse sound speed \cite{LL}. The parameters $\omega_0$ and $\gamma$
are the oscillator eigenfrequency in the absence of friction and the damping decrement,
respectively. Note that when a solid medium surrounds the nanoparticle, we must set
${\cal{E}}_{\text s}=0$ in expressions~(\ref{eq 6}). However, if the nanoparticle is embedded in
a liquid medium, then only the second terms in expressions~(\ref{eq 6}) must be retained.
The quantity $1/\gamma$ specifies the time over which the oscillations are completely
damped. It depends on the nanoparticle radius \cite{HBP}.

Equation~(\ref{eq 5}) formally describes the motion of the oscillator under the action
of an external driving force of the form $\delta p(t)R/\rho$.
We will consider that there are no oscillator eigenmodes without an
external force, and that they are generated only by an external force.
If the driving force acting on the oscillator,
for example, is $\delta$-shaped 
\be
 \label{eq 7}
  \delta p(t)= p_{\text{im}} \delta(t-\tau_0),
   \ee
then the solution of equation~(\ref{eq 5}) with the initial conditions $\phi=0$
and $\phi' = - Rp_{\text{im}}/\rho$ for $t = \tau_0$ is given by
\be
 \label{eq 8}
  \phi(t)= - \re^{-\gamma(t-\tau_0)}
   \sin\left[\omega(t-\tau_0)\right]R p_{\text{im}}/{\rho\omega},
    \ee
where $\omega=\sqrt{\omega^2_0-\gamma^2_0}$, provided that $\omega_0 > \gamma$.

When the time dependence of the function $T_{\text e}(t)$ is known, the value of
$p_{\text{im}}$ can be determined explicitly. Comparing expressions~(\ref{eq 4})
and (\ref{eq 7}), we find
\be
 \label{eq 9}
  \alpha\int_0^{\infty}T^2_{\text e}(t-\tau_0) \,\rd t = p_{\text{im}}\,.
   \ee
The variation of the electron temperature with time can be modelled
or determined analytically.

Let us suppose that initially (at $t = \tau_0$) the oscillator
is not displaced, i.e., is at rest, and obeys the initial conditions $\phi(\tau_0) = 0$
and $\phi'(\tau_0)=0$. Then, the solution of equation~(\ref{eq 5}) for an arbitrary
form of the driving force can be presented in an integral form \cite{Kam}
\be
 \label{eq 10}
  \phi(t)= -\frac{R \re^{-\gamma t}}{\rho\omega}\int_{\tau_0}^{t}
   \delta p(\tau)\,\re^{\gamma\tau}\sin\left[\omega(t-\tau)\right] \rd\tau,
    \ee
with the conditions that $\delta p(\tau) \neq 0$ and $t > 0$.

Equation~(\ref{eq 10}) easily answers the question of how the oscillator behaves
after an excess pressure $\delta p(\tau)$ has acted on it over a short
time interval $(\tau_0, \tau_1)$.
The upper limit in equation~(\ref{eq 10}) for this case, evidently, will be
the quantity $\tau_1$. Then, using the mean-value theorem in equation~(\ref{eq 10}),
we obtain
\be
 \label{eq 11}
  \phi(t)= -R I \re^{-\gamma t} \re^{\gamma\xi \tau_1}
   \sin(\omega t)/{\rho\omega},
    \ee
where $\tau_0/\tau_1 < \xi < 1$, while $\int_{\tau_0}^{\tau_1}
\delta p(\tau) \rd\tau \equiv I$ presents the impulse of the excess pressure of the electron gas.
If $\omega_0 \tau_1\xi \ll 2\piup$, then the explicit form of $\delta p(\tau)$ is not essential;
an important point is only the value of~$I$. When $\omega_0\gg\gamma$, the maximum
displacement $u_{\text{max}}$ caused by the excess pressure ``impact'', with using potential
method and equation~(\ref{eq 11}), will be
\be
 \label{eq 12}
  u_{\text{max}} = I/{\rho\omega_0 R}.
    \ee
Taking into account here equation~(\ref{eq 6}) for the case ${\cal{E}}_{\text s} = 0$, we obtain
\be
 \label{eq 13}
  u_{\text{max},\,{\cal{E}}_{\text s}=0} = I/{2\sqrt{\mu\rho}},
   \ee
i.e., the magnitude of $u_{\text{max},\,{\cal{E}}_{\text s}=0}$ is independent of the particle radius
(and is more independent since the rigidity modulus and mass density of the nanoparticle
are smaller). On the other hand, if the ${\cal{E}}_{\text s} > 0$, then a dependence on the
particle radius shows up, and the magnitude of $u_{\text{max}}$ is decreased as $R$ is reduced.
The maximum displacement over the time $\tau_1-\tau_0$ with account for the finite damping,
has the form:
\be
 \label{eq 14}
  u_{\text{max},\,{\cal{E}}_{\text s}=0} = \frac{I(1-R\gamma/s_{\text L})}{2\sqrt{\mu\rho}
   \sqrt{1-(s_{\text T}/s_{\text L})^2}} \re^{-\gamma(\tau_1-\tau_0)}.
    \ee
Therefore, it is possible to determine the pressure impulse $I$
by measuring the maximum displacement of the physical object.

\section{ Power radiated by nanoparticle surface}
A certain part of the accumulated by MN's energy is carried away
from the particle into the surrounding matrix by sound waves. The instantaneous elastic
energy flux from the surface of the sphere (for the case of ${\cal{E}}_{\text s} = 0$)
can be calculated using a formula \cite{FH}
\be
 \label{eq 15}
  {\cal W}(t)= S\,\delta p(t)\,\upsilon_r(R,t).
   \ee
Here, $S$ is the area of the particle surface and $\upsilon_r$ is the radial
or vibrational velocity.

In terms of time, conventionally, two stages in this process can be identified:
one is an impetuous rise in $\delta p$ during an ultrashort
time spell ($t\sim\tau_0$) and another one is a slower decrease (under $t > \tau_0$)
owing to a transfer of the electron energy to the surrounding matrix. The dependence
of the $\delta p(t)$ within the second time spell,
can be well described by expression~(\ref{eq 4}).

In order to take an explicit account of the behavior of the pressure in the initial
time spell, the function $\delta p(t)$ can be presented as the product of two functions,
\be
 \label{eq 16}
  \delta p(t)\rightarrow\theta(t-\tau_0)\cdot\alpha T_{\text e}^2(t),
   \ee
where we use equation~(\ref{eq 4}), while
$\theta(t-\tau_0)$ is the Heaviside unit-step function that specifies
the pressure behavior when $t\rightarrow\tau_0$.
Let us also suppose  that the electron temperature varies with time
in accordance with the rule
\be
 \label{eq 17}
  T_{\text e}(t)= T_{\text e}(\tau_0)\, \re^{-\beta(t-\tau_0)}, \qquad \beta = \frac{g_R}{C_{\text e}}\,,
   \qquad C_{\text e} = 3\alpha T_{\text e}\,,
    \ee
where $C_{\text e}\equiv C_{\text e}\big(T_{\text e}(\tau_0)\big)$ is the heat capacity of the electron gas,
which depends on the electron temperature taken at the instant of time $t = \tau_0$.
The electron-phonon coupling constant in
equation~(\ref{eq 17}) can be computed from the formula
\be
 \label{eq 18}
  g_R=\frac{27}{16}\frac{n}{m}\frac{\alpha}{k_{\text B}}\frac{1}{\rho_p R}
   \left(\frac{\piup\hbar}{a}\right)^3\left(\frac{U_1}{A}\right)^2.
    \ee
Here, $a$ is the lattice constant for the MN, $\rho_p$ denotes the mass density
of the MN, $U_1$ is the energy required to detatch the first electron from the
neutral unexcited atom, and $A$ refers to the electron work function of the metal.
Formula (\ref{eq 18}) follows from equation (115) of the work \cite{FNT}
using the equation~(\ref{eq 4}) and the following relationships
\be
 \label{eq 181}
  \omega_{\text D}\simeq\sqrt{\frac{\sigma}{\rho_p}\left(\frac{\piup}{a}\right)^3},
  \qquad \upsilon_{\text F}= \frac{\hbar}{m} \big(3n\piup^2\big)^{1/3},
    \ee
where $\omega_{\text D}$ is the Debye frequency and $\sigma$ is the surface energy density.

Taking into account equations~(\ref{eq 16}) and (\ref{eq 17}),
after integrating with respect to $\tau$ within limits $0<\tau<t$
and some transformations, we finally obtain
\begin{align}
 \label{eq 19}
  {\cal W}(t)&= \frac{4\piup R^2}{\rho s_{\text L}}\frac{\delta p(t)\,
   \delta p_m(t)}{\omega^2_0+4\beta(\beta-\gamma)}\bigg(-2\beta
    \left(\frac{s_{\text L}}{R}-2\beta\right) + 
     \re^{(2\beta-\gamma)t_0}\bigg\{\left[\omega^2_0+2\beta
      \left(\frac{s_{\text L}}{R}-2\gamma\right)\right]
       \cos\left(\omega t_0\right)\nonumber \\&\quad + 
        \left[(\omega^2_0-2\beta\gamma)\left(\frac{s_{\text L}}{R}-\gamma\right)-2\beta\omega^2\right]
          \frac{\sin\left(\omega t_0\right)}{\omega}\bigg\}\bigg),
            \end{align}
where $t_0=t-\tau_0$. Equation~(\ref{eq 19}) gives the energy per unit time carried
out by spherical sound waves from the nanoparticle into the surrounding medium.

The analysis show that the power of the sonic signal ${\cal W}(t)$ is of
the form of damped oscillations. The number of oscillations
depends much on the density of the matrix material.
The amplitude of the sound power is increased with
an increase in the radius of MN, and it is fully damped over longer times.
For different metals (but with the same radius) the amplitude of ${\cal W}(t)$ will
be greater for MNs with a higher coupling constant $g_R$.

The maximum power of the acoustic signal is reached at the time instant
$t_0 = 0$. From equation~(\ref{eq 19}), it is equal to
\be
 \label{eq 20}
  {\cal W}_{\text{max}}\big|_{t_0=0}= 4\piup R^2[\delta p(\tau_0)]^2/\rho s_{\text L}.
   \ee

The temperature decrease of the electron gas
is proportional to the product $\beta T_{\text e}$,
which characterizes the cooling rate of the electron gas. Since the fall of the electron
temperature occurs much more slowly than the electron's heating, we can assume
that the influence of this mechanism on the sound power is negligible in the
first approximation. Formally, this allows us to direct $\beta\rightarrow 0$
in equation~(\ref{eq 19}). In this case, the sound signal will 
correspond to exponentially damped oscillations.

The time dependence of the excess pressure in the case of
$\beta\rightarrow 0$, can also be estimated using the equation
\be
 \label{eq 21}
  \delta p(t)\big|_{\beta\rightarrow 0}\approx -2g_R \,t\, T_{\text e}(\tau_0)/3,
   \ee
which follows from the energy balance equation for the electrons.
The maximum power of the signal [equation~(\ref{eq 20})], with the account
for approximation~(\ref{eq 21}), is
\be
 \label{eq 22}
  {\cal W}_{\text{max}}\big|_{t=\tau_0,\, \beta\rightarrow 0}= \frac{16}{9}
   \frac{\piup}{\rho s_{\text L}} [g_R \,\tau_0 \,R \,T_{\text e}(\tau_0)]^2,
    \ee
where the dependence on the laser pulse duration appears explicitly.

\section{Total energy of the oscillations of the metal nanoparticle}

We now determine the total energy transferred from the pulsating spherical
surface to the surrounding matrix. We will consider the case $\beta\neq 0$,
when the power ${\cal W}(t)$ is given by equation~(\ref{eq 19}). Integrating equation~(\ref{eq 19})
with respect to time over the interval $0 < t < \infty$, in view of
expressions~(\ref{eq 4}) and (\ref{eq 6}) at ${\cal{E}}_{\text s}=0$, we obtain
\be
 \label{eq 23}
  E = \int_{\tau_0}^{\infty} {\cal W}(t)\,\rd t =
   \frac{2\piup R}{\rho s_{\text L}}
    \frac{s_{\text L}+2\beta R}{\omega^2_0+4\beta(\beta+\gamma)}\alpha^2 T^4_{\text e}.
     \ee
Equation~(\ref{eq 23}) shows that the total energy is proportional to the fourth power
of the electron temperature. In the case $\beta = 0$, it is easy to reveal that
this energy depends on the volume of the particle and is reciprocal to the rigidity
modulus of the medium.

On the other hand, the work necessary to shift the spherical surface
at the maximal distance $u_{\text{max}}$ [defined by equation~(\ref{eq 14})], is
\be
 \label{eq 24}
  E = 8\piup R u_{\text{max}}^2\alpha T^2_{\text e}.
    \ee
Comparing it with equation~(\ref{eq 23}), one finds
\be
 \label{eq 25}
  T_{\text e} = 2u_{\text{max}}\sqrt{\frac{\rho s_{\text L}}{\alpha}
   \frac{\omega^2_0+4\beta(\beta+\gamma)}{s_{\text L}+2\beta R}}\,.
    \ee

\begin{figure}[!t]
\centering\includegraphics[width=0.5\textwidth]{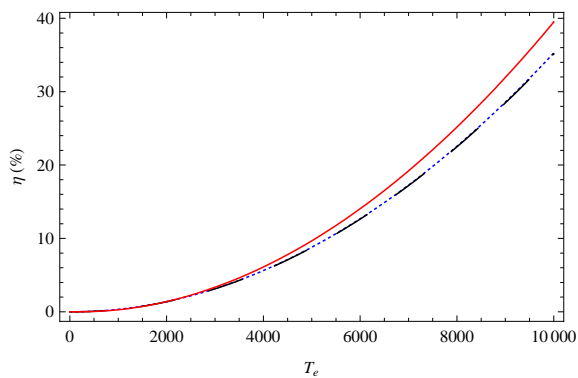}
\caption{(Color online) Dependence of the sound signal power efficiency on electron
temperature for Cu nanoparticle (solid curve), Ag (dashed curve), and Au (dotted curve),
embedded in plexiglass.}
\label{fig:2}
\end{figure}

Another way to determine $T_{\text e}$ is as follows.
The total energy obtained by the electron gas in the MN is
\be
 \label{eq 26}
  E_{\text{tot}} = 3 V \alpha T^2_{\text e}/2.
    \ee
The efficiency of the energy transfer $\eta$ from the pulsating spherical MN
to the sound oscillations can be estimated taking the ratio of equations~(\ref{eq 23})
and (\ref{eq 26}). It has the simplest form in the case $\beta\rightarrow 0$
\be
 \label{eq 27}
  \eta(T)\equiv E|_{\beta\rightarrow 0}/E_{\text{tot}} = \alpha T^2_{\text e}/4\mu.
    \ee
The temperature dependence of $\eta(T)$ for noble nanoparticles embedded in
a plexiglass matrix is plotted in figure~\ref{fig:2}. The calculations were done using
equations~(\ref{eq 23}) and (\ref{eq 26}), with account for the characteristics
both of the matrix and the MNs given in tables~\ref{tab:1} and \ref{tab2}. The efficiency
increases as the square of the temperature. Actually, it increases if the rigidity modulus of the medium in which the nanoparticle is embedded gets smaller.
So, knowing from experiment the efficiency $\eta$, it is easy to estimate
the electron temperature in MN.
The highest possible $\eta_T$ for noble metal nanoparticles with $R=100$~{\AA}
embedded in a plexiglass matrix are presented in  table~\ref{tab2} at $T_{\text e} =  10^4$.

\begin{table}[!t]
\begin{center}
\caption{Constants for the plexiglass \cite{Tab}.}
\vspace{2ex}
\label{tab:1}
\begin{tabular}{|c|c|c|c|c|c|}
\hline\hline 
\multirow{2}{*}{Medium} & $K$ & $\mu$ & $\rho$ &$s_{\text L}$ & $s_{\text T}$ \\
&(dyn/cm$^2$)&(dyn/cm$^2$)&(g/cm$^3)$&(cm/s)&(cm/s)\\
\hline\hline
Plexiglass
&5.83$\cdot 10^{10\strut}$&1.48$\cdot 10^{10}$& 1.18 & 2.57 $\cdot 10^{5}$&1.12$\cdot 10^{5}$\\
\hline\hline
\end{tabular}
\end{center}
\end{table}

\begin{table}[!t]
\begin{center}
\caption{Physical parameters of the noble MNs.}
\vspace{2ex}
\label{tab2}
\begin{tabular}{|c|c|c|c|c|c|}
\hline\hline
 Metal
 &$l_{\text D}$ (\AA) &$\alpha$ $\left(\frac{\text{erg}}{\text{cm}^3~\text{K}^2}\right)$&$n$ \big(cm$^{-3}$\big) \cite{Kit}&$\eta_{10^4}$ (\%)\\
 \hline
 \hline
 Cu& $ 1197 $&$ 235.5  $&$8.45\cdot 10^{22\strut}$&$ 39.5$\\\hline
 Ag& $ 1552 $&$ 208.35 $&$5.85\cdot 10^{22\strut}$&$35.15$\\\hline
 Au& $ 1968 $&$ 208.9  $&$5.90\cdot 10^{22\strut}$&$35.3 $\\\hline\hline
\end{tabular}
 \end{center}
\end{table}

\section{Conclusions} 
The method has been proposed for the estimation of hot electron
temperature in metallic nanoparticles embedded in a dielectric matrix under irradiation
of ultrashort laser pulses. Analytic expressions have been derived for the amplitude
and power of  longitudinal spherical sound oscillations as function of the
density and elastic properties of the medium, the laser pulse duration, electron
temperature, nanoparticle radius, and electron-phonon coupling constants.

The maximal displacement of any oscillator in the MN's
material is due to the ``impact'' of the excess pressure of the electron gas.
The magnitude of the sound signal power at the moment of the laser pulse
ending can be used to estimate the maximal electron temperature in the MN.
The latter is determined by the cooling rate of the
electron gas and the rate of its pressure change.

The efficiency with which
the energy of the hot electrons is carried away by sound oscillations has been
examined for the noble metals. It has been shown that the sound energy transfer
efficiency is considerably higher in the medium with smaller rigidity moduli.

\section*{Acknowledgements}
Author is indebted to DFFD of Ukraine for  single financial support of this work.

\ukrainianpart

\title{Визначення температури електронів у підігрітій металевій наночастинці}
\author{М.І. Григорчук}
\address{
Інститут теоретичної фізики ім.~М.М.~Боголюбова НАН України, \\
 вул.~Метрологічна, 14-б, 03680 Київ, Україна 
}

\makeukrtitle

\begin{abstract}
\tolerance=3000%
Запропоновано метод визначення температури гарячих електронів в металевій
наночастинці, що знаходиться в середовищі під дією ультракоротких лазерних
імпульсів. Одержані амплітуда і потужність поздовжних сферичних акустичних
коливань як функція густини і пружних властивостей середовища, тривалості
лазерного імпульсу, радіусу частинки, електрон-фононної константи зв'язку та
електронної температури. Зроблено оцінку ефективності передачі енергії електронів
від підігрітих благородних наночастинок в оточуюче їх середовище за різних
температур електронів.
\keywords електронна температура, металева наночастинка, ультракороткі лазерні імпульси

\end{abstract}

\end{document}